\def\eck#1{\left\lbrack #1 \right\rbrack}
\def\eckk#1{\bigl[ #1 \bigr]}
\def\rund#1{\left( #1 \right)}
\def\abs#1{\left\vert #1 \right\vert}

\def\ave#1{\left\langle #1 \right\rangle}

\def\part#1#2{{\partial #1\over\partial #2}}

\def\Re{{\cal R}\hbox{e}}

\def\A{{\cal A}}

\def\d{{\rm d}}

\def\eps{{\epsilon}}

\def\vp{\varphi}
\def\vt{{\vartheta}}

\def\Real{{\rm I\mathchoice{\kern-0.70mm}{\kern-0.70mm}{\kern-0.65mm}%
  {\kern-0.50mm}R}}
  % Symbol fuer reelle Zahlen.                                 MJL
\def\C{\rm C\kern-.42em\vrule width.03em height.58em depth-.02em
       \kern.4em}
\font \bolditalics = cmmib10
\def\bx#1{\leavevmode\thinspace\hbox{\vrule\vtop{\vbox{\hrule\kern1pt
        \hbox{\vphantom{\tt/}\thinspace{\bf#1}\thinspace}}
      \kern1pt\hrule}\vrule}\thinspace}

\def \vc #1{{\textfont1=\bolditalics \hbox{$\bf#1$}}}
{\catcode`\@=11
\gdef\SchlangeUnter#1#2{\lower2pt\vbox{\baselineskip 0pt \lineskip0pt
  \ialign{$\m@th#1\hfil##\hfil$\crcr#2\crcr\sim\crcr}}}
  % kopiert von \@vereq aus dem TeXbook, Seite 360.
}

\def\ueber#1#2{{\setbox0=\hbox{$#1$}%
  \setbox1=\hbox to\wd0{\hss$\scriptscriptstyle #2$\hss}%
  \offinterlineskip
  \vbox{\box1\kern0.4mm\box0}}{}}

\def\bx#1{\leavevmode\thinspace\hbox{\vrule\vtop{\vbox{\hrule\kern1pt
        \hbox{\vphantom{\tt/}\thinspace{\bf#1}\thinspace}}
      \kern1pt\hrule}\vrule}\thinspace}

\def\SFB{{This work was supported by the ``Sonderforschungsbereich
375-95 f\"ur
Astro--Teil\-chen\-phy\-sik" der Deutschen For\-schungs\-ge\-mein\-schaft.}}
%\input mssymb
%\input sp3
 
% Kodierungen zur automatischen Erstellung des Layouts:
\magnification=\magstep1
\input epsf
\voffset= 0.0 true cm
\vsize=19.8 cm     % wird im Ausdruck 23.7
\hsize=13.5 cm
\hfuzz=2pt
\tolerance=500
\abovedisplayskip=3 mm plus6pt minus 4pt
\belowdisplayskip=3 mm plus6pt minus 4pt
\abovedisplayshortskip=0mm plus6pt
\belowdisplayshortskip=2 mm plus4pt minus 4pt
\predisplaypenalty=0
\footline={\tenrm\ifodd\pageno\hfil\folio\else\folio\hfil\fi}
%-----------------------------------------------------------------------

\def\la{\mathrel{\hbox{\rlap{\hbox{\lower4pt\hbox{$\sim$}}}\hbox{$<$}}}}
\def\ga{\mathrel{\hbox{\rlap{\hbox{\lower4pt\hbox{$\sim$}}}\hbox{$>$}}}}

\def\utw{\smash{\rlap{\lower5pt\hbox{$\sim$}}}}
\def\udtw{\smash{\rlap{\lower6pt\hbox{$\approx$}}}}

\def\getsto{\mathrel{\hbox{\rlap{$\gets$}\hbox{\raise2pt\hbox{$\to$}}}}}
\def\lid{\mathrel{\hbox{\rlap{\hbox{\lower4pt\hbox{$=$}}}\hbox{$<$}}}}
\def\gid{\mathrel{\hbox{\rlap{\hbox{\lower4pt\hbox{$=$}}}\hbox{$>$}}}}
\def\sol{\mathrel{\hbox{\rlap{\hbox{\raise4pt\hbox{$\sim$}}}\hbox{$<$}}}
}
\def\sog{\mathrel{\hbox{\rlap{\hbox{\raise4pt\hbox{$\sim$}}}\hbox{$>$}}}
}
\def\lse{\mathrel{\hbox{\rlap{\hbox{\raise4pt\hbox{$<$}}}\hbox{$\simeq$}
}}}
\def\gse{\mathrel{\hbox{\rlap{\hbox{\raise4pt\hbox{$>$}}}\hbox{$\simeq$}
}}}
\def\grole{\mathrel{\hbox{\lower2pt\hbox{$<$}}\kern-8pt
\hbox{\raise2pt\hbox{$>$}}}}
\def\leogr{\mathrel{\hbox{\lower2pt\hbox{$>$}}\kern-8pt
\hbox{\raise2pt\hbox{$<$}}}}
\def\loa{\mathrel{\hbox{\rlap{\hbox{\lower4pt\hbox{$\approx$}}}\hbox{$<$
}}}}
\def\goa{\mathrel{\hbox{\rlap{\hbox{\lower4pt\hbox{$\approx$}}}\hbox{$>$
}}}}

%-----------------------------------------------------------------------
%
%  Fontdefinitionen
%

% vektor-fonts
%\font\halbcurs gibt es schon als \tams=cmmib10
\font\kleinhalbcurs=cmmib10 scaled 833
% petit-fonts
\font\eightrm=cmr8
\font\sixrm=cmr6
\font\eighti=cmmi8
\font\sixi=cmmi6
\skewchar\eighti='177 \skewchar\sixi='177
\font\eightsy=cmsy8
\font\sixsy=cmsy6
\skewchar\eightsy='60 \skewchar\sixsy='60
\font\eightbf=cmbx8
\font\sixbf=cmbx6
\font\eighttt=cmtt8
\hyphenchar\eighttt=-1
\font\eightsl=cmsl8
\font\eightit=cmti8

\font\bxf=cmbx10
%-------------------------------------------------------------------
% Definition der versal griechischen Buchstaben
%=======================================================================
  \mathchardef\Gamma="0100
  \mathchardef\Delta="0101
  \mathchardef\Theta="0102
  \mathchardef\Lambda="0103
  \mathchardef\Xi="0104
  \mathchardef\Pi="0105
  \mathchardef\Sigma="0106
  \mathchardef\Upsilon="0107
  \mathchardef\Phi="0108
  \mathchardef\Psi="0109
  \mathchardef\Omega="010A
%-----------------------------------------------------------------------
\def\rahmen#1{\vskip#1truecm}
% Abbildungen
\def\begfig#1cm#2\endfig{\par
\setbox1=\vbox{\rahmen{#1}#2}%
\dimen0=\ht1\advance\dimen0by\dp1\advance\dimen0by5\baselineskip
\advance\dimen0by0.4true cm
\ifdim\dimen0>\vsize\pageinsert\box1\vfill\endinsert
\else%keine seitenhohe Abbildung
\dimen0=\pagetotal\ifdim\dimen0<\pagegoal
\advance\dimen0by\ht1\advance\dimen0by\dp1\advance\dimen0by1.4true cm
\ifdim\dimen0>\vsize
\topinsert\box1\endinsert
\else\vskip1true cm\box1\vskip4true mm\fi
\else\vskip1true cm\box1\vskip4true mm\fi\fi}
%-------------------------------------------------------------------
% Abbildungslegenden
% Falls Text kleiner als eine volle Zeile, zentriert.
\def\figure#1#2{\smallskip\setbox0=\vbox{\noindent\petit{\bf Fig.\ts#1.\
}\ignorespaces #2\smallskip
\count255=0\global\advance\count255by\prevgraf}%
\ifnum\count255>1\box0\else
\centerline{\petit{\bf Fig.\ts#1.\ }\ignorespaces#2}\smallskip\fi}
%-----------------------------------------------------------------

%-----------------------------------------------------------------
% Tabellenkoepfe

%-------------------------------------------------------------------
\def\begtab#1cm#2\endtab{\par
\ifvoid\topins\midinsert\vbox{#2\rahmen{#1}}\endinsert
\else\topinsert\vbox{#2\kern#1true cm}\endinsert\fi}
\def\rahmen#1{\vskip#1truecm}
%-----------------------------------------------------------------
\def\begpet{\vskip6pt\bgroup\petit}
\def\endpet{\vskip6pt\egroup}
% Referenzen
\def\begref{\par\bgroup\petit
\let\it=\rm\let\bf=\rm\let\sl=\rm\let\INS=N}
% Jede Referenz bleibt komplett, kein Seitenumbruch dazwischen
%\def\ref#1{\filbreak\if N\INS\let\INS=Y\vbox{\sec{References}}
%\fi\hangindent\parindent
%\hangafter=1\noindent\hbox to\parindent{#1\hfil}\ignorespaces}
%\let\endref=\endpet% Ende der Referenzen
%-------------------------------------------------------------------
%\def\vec#1{\hbox{\textfont1=\tamss\scriptfont1=\kleinhalbcurs
%\textfont0=\bxf\scriptfont0=\sevenbf
%$#1$}}
%---------------------------------------------------------------------
\def\petit{\def\rm{\fam0\eightrm}%
\textfont0=\eightrm \scriptfont0=\sixrm \scriptscriptfont0=\fiverm
 \textfont1=\eighti \scriptfont1=\sixi \scriptscriptfont1=\fivei
 \textfont2=\eightsy \scriptfont2=\sixsy \scriptscriptfont2=\fivesy
 \def\it{\fam\itfam\eightit}%
 \textfont\itfam=\eightit
 \def\sl{\fam\slfam\eightsl}%
 \textfont\slfam=\eightsl
 \def\bf{\fam\bffam\eightbf}%
 \textfont\bffam=\eightbf \scriptfont\bffam=\sixbf
 \scriptscriptfont\bffam=\fivebf
 \def\tt{\fam\ttfam\eighttt}%
 \textfont\ttfam=\eighttt
 \normalbaselineskip=9pt
 \setbox\strutbox=\hbox{\vrule height7pt depth2pt width0pt}%
 \normalbaselines\rm
\def\vec##1{\setbox0=\hbox{$##1$}\hbox{\hbox
to0pt{\copy0\hss}\kern0.45pt\box0}}}%
\let\ts=\thinspace
%-------------------------------------------------------------------
%Fonts fur die Uberschriften:
%
\font \tafontt=     cmbx10 scaled\magstep2
\font \tafonts=     cmbx7  scaled\magstep2
\font \tafontss=     cmbx5  scaled\magstep2
\font \tamt= cmmib10 scaled\magstep2
\font \tams= cmmib10 scaled\magstep1
\font \tamss= cmmib10
\font \tast= cmsy10 scaled\magstep2
\font \tass= cmsy7  scaled\magstep2
\font \tasss= cmsy5  scaled\magstep2
\font \tasyt= cmex10 scaled\magstep2
\font \tasys= cmex10 scaled\magstep1
\font \tbfontt=     cmbx10 scaled\magstep1
\font \tbfonts=     cmbx7  scaled\magstep1
\font \tbfontss=     cmbx5  scaled\magstep1
\font \tbst= cmsy10 scaled\magstep1
\font \tbss= cmsy7  scaled\magstep1
\font \tbsss= cmsy5  scaled\magstep1

%-----------------------------------------------------------------
\newbox\chsta\newbox\chstb\newbox\chstc
\def\centerpar#1{{\advance\hsize by-2\parindent
\rightskip=0pt plus 4em
\leftskip=0pt plus 4em
\parindent=0pt\setbox\chsta=\vbox{#1}%
\global\setbox\chstb=\vbox{\unvbox\chsta
\setbox\chstc=\lastbox
\line{\hfill\unhbox\chstc\unskip\unskip\unpenalty\hfill}}}%
\leftline{\kern\parindent\box\chstb}}
%---------------------------------------------------------------
 % Beginn Ueberschrift 1. Ordnung
 \def \chap#1{%\goodbreak
    \vskip24pt plus 6pt minus 4pt
    \bgroup
 \textfont0=\tafontt \scriptfont0=\tafonts \scriptscriptfont0=\tafontss
 \textfont1=\tamt \scriptfont1=\tams \scriptscriptfont1=\tamss
 \textfont2=\tast \scriptfont2=\tass \scriptscriptfont2=\tasss
 \textfont3=\tasyt \scriptfont3=\tasys \scriptscriptfont3=\tenex
     \baselineskip=18pt
     \lineskip=18pt
     \raggedright
     \pretolerance=10000
     \noindent
     \tafontt
     \ignorespaces#1\vskip7true mm plus6pt minus 4pt
     \egroup\noindent\ignorespaces}%
%------------------------------------------------------
 % Beginn Ueberschrift 2. Ordnung
 \def \sec#1{%\goodbreak
     \vskip25true pt plus4pt minus4pt
     \bgroup
 \textfont0=\tbfontt \scriptfont0=\tbfonts \scriptscriptfont0=\tbfontss
 \textfont1=\tams \scriptfont1=\tamss \scriptscriptfont1=\kleinhalbcurs
 \textfont2=\tbst \scriptfont2=\tbss \scriptscriptfont2=\tbsss
 \textfont3=\tasys \scriptfont3=\tenex \scriptscriptfont3=\tenex
     \baselineskip=16pt
     \lineskip=16pt
     \raggedright
     \pretolerance=10000
     \noindent
     \tbfontt
     \ignorespaces #1
     \vskip12true pt plus4pt minus4pt\egroup\noindent\ignorespaces}%
%------------------------------------------------------
 % Beginn Ueberschrift 3. Ordnung
 \def \subs#1{%\goodbreak
     \vskip15true pt plus 4pt minus4pt
     \bgroup
     \bxf
     \noindent
     \raggedright
     \pretolerance=10000
     \ignorespaces #1
     \vskip6true pt plus4pt minus4pt\egroup
     \noindent\ignorespaces}%
%------------------------------------------------------
 % Beginn Ueberschrift 4. Ordnung
 \def \subsubs#1{%\goodbreak
     \vskip15true pt plus 4pt minus 4pt
     \bgroup
     \bf
     \noindent
     \ignorespaces #1\unskip.\ \egroup
     \ignorespaces}
%-------------------------------------------------------------------
\def\footnoterule{\kern-3pt\hrule width 2true cm\kern2.6pt}
% Fussnoten-macros
\newcount\footcount \footcount=0
\def\advftncnt{\advance\footcount by1\global\footcount=\footcount}
% Automatisch numerierte Fussnote, Fussnotentex in petit
\def\fonote#1{\advftncnt$^{\the\footcount}$\begingroup\petit
       \def\textindent##1{\hang\noindent\hbox
       to\parindent{##1\hss}\ignorespaces}%
\vfootnote{$^{\the\footcount}$}{#1}\endgroup}
%-------------------------------------------------------------------
% Acknowledgement

%-------------------------------------------------------------------
% Satz fur bye:
\newcount\sterne
\outer\def\byebye{\bigskip\typeset
\sterne=1\ifx\speciali\undefined\else
\bigskip Special caracters created by the author
\loop\smallskip\noindent special character No\number\sterne:
\csname special\romannumeral\sterne\endcsname
\advance\sterne by 1\global\sterne=\sterne
\ifnum\sterne<11\repeat\fi
\vfill\supereject\end}
\def\typeset{\centerline{\petit This article was processed by the author
using the \TeX\ Macropackage from Springer-Verlag.}}
 
%**end of header
%\pageno=0
%scriptum
%\vfill\eject
\voffset=0pt

\baselineskip=14pt

\def\m{M_{\rm ap}}
\def\N{{\cal N}}
\def\s{{({\rm s})}}
\chap{\centerline{Cosmic shear and biasing}}
\centerline{\bf Peter Schneider}
\medskip
\centerline{\bf Max-Planck-Institut f\"ur Astrophysik}
\centerline{\bf Postfach 1523}
\centerline{\bf D-85740 Garching, Germany}
\bigskip
\sec{Abstract}
The correlation between cosmic shear as measured by the image
distortion of high-redshift galaxies and the number counts of
foreground galaxies is calculated. For a given power spectrum of the
cosmic density fluctuations, this correlation is proportional to the
bias factor, which can thus directly be measured. In addition, this
correlation provides a first-order measure of cosmic shear and is
therefore easier to observe than quadratic measures hitherto
proposed. Analytic approximations show that the expected
signal-to-noise ratio of the correlation is large, so that a
significant detection is possible with a moderate amount of data; in
particular, it is predicted that the ongoing ESO Imaging Survey (EIS)
will be able to detect this correlation on scales of $\sim 10'$ at a
3-$\sigma$ level, and at with higher significance on smaller
angular scales.

\sec{1 Introduction}
The distortion of high-redshift galaxy images by the (tidal)
gravitational field of intervening matter inhomogeneities (often
called `cosmic shear') can be used
to study the intervening mass distribution. In particular, if the
large-scale structure of the (dark) matter is considered, the
observable image distortions constrain the statistical properties of
the cosmic matter distribution. This method of determining
the power spectrum of cosmic density fluctuations has been
investigated recently in considerable 
detail (Blandford et al.\ 1991;  Miralda-Escud\'e 1991; Kaiser 1992,
1996, hereafter K92, K96, respectively; Villumsen 1996;
van Waerbeke, Bernardeau \&
Mellier 1997; Jain \& Seljak 1997; Sanz, Mart\'\i nez-Gonz\'alez \&
Ben\'\i tez 1997; Schneider et al.\ 1997a, hereafter SvWJK). A first
significant detection of 
cosmic shear has been reported in Schneider et al.\ (1997b).

Cosmic shear probes the statistical properties of the projected density
fluctuations, where the projection takes into account
the redshift distribution of the source galaxies and geometric
factors. Similarily, the surface number density of galaxies is
obtained by a projection of the three-dimensional galaxy
distribution. Provided galaxies trace the underlying dark matter
distribution, these two projected fields are correlated. 

In this letter, this correlation is investigated, aiming at a
method to constrain the bias factor $b$ which relates the number
density fluctuations of galaxies to those of the underlying dark
matter distribution.\fonote{A different method to obtain the bias
factor directly lensing uses the magnification effect which causes a
correlation between foreground galaxies and high-redshift QSOs (e.g.,
Bartelmann 1995, Sanz et al.\ts 1997, Dolag \& Bartelmann 1997) and
changes the angular two-point correlation function of high-redshift
galaxies (e.g., Moessner, Jain \& Villumsen 1997).}
A brief summary of the aperture mass statistics
as a measure for cosmic shear
is presented in Sect.\ts 2, and an analogous statistics is introduced
for the galaxy number counts. The general expression for the
correlation between these two measures is derived in Sect.\ts 3, and
practical estimators are considered in Sect.\ts 4. Focusing on large
angular scales, linear theory presents a
useful approximation for the growth of cosmic density fluctuations; in
Sect.\ts 5,
the general expressions will be evaluated in
this approximation. In particular, it is shown that the
signal-to-noise ratio for the correlation coefficient is of order
unity even in a single field. Therefore, this correlation should be
easily detectable in currently conducted wide-field surveys, such as
the EIS (Renzini \& da Costa 1997).
In particular, the significant verification of
this correlation is probably the easiest way to detect cosmic shear.

\sec{2 Aperture mass and number counts}
In this section we briefly summarize the $\m$-statistics for cosmic
shear, and introduce a similar statistics for the number counts of
(foreground) galaxies, following the notation of SvWJK. 

Light propagation through a slightly inhomogeneous Universe can be
described by an equivalent single-plane gravitational lens
equation, to first order in the Newtonian gravitational potential (see
SvWJK for a detailed discussion of this point). For sources with a
redshift probability density $p_z(z)\,\d z=p_w(w)\,\d w$, where $w$ is
the comoving distance out to redshift $z$, the dimensionless surface
mass density at angular position $\vc\theta$ of this single-plane lens is
$$
\kappa(\vc\theta)%:=\int \d w\; p_w(w)\,\kappa(\vc\theta,w) =
={3\over 2}\rund{H_0\over c}^2\Omega_{\rm d}
\int_0^{w_{\rm H}} \d w\;
g(w)\,f_K(w){\delta\rund{f_K(w)\vc\theta,w}\over a(w)}\; ,
\eqno (1)
$$
where $\delta$ is the density contrast, $a=(1+z)^{-1}$ the cosmic expansion
factor, $\Omega_{\rm d}$ the  density
parameter in dust at present, 
$
g(w):=\int_w^{w_{\rm H}}\d w'\;p_w(w')\*{f_K(w'-w) / f_K(w')}
$
is the source-averaged distance ratio $D_{\rm ds}/D_{\rm s}$ for a
density fluctuation at distance $w$, $f_K(w)$ is the
comoving angular diameter distiance to comoving distance $w$,
and $w_{\rm H}$ is the comoving
distance to the horizon.

The Jacobi matrix, which describes the locally linearized lens mapping,
reads $\A_{ij}(\vc\theta)=\delta_{ij}-\psi_{,ij}(\vc\theta)$, where
indices preceded by a comma denote partial derivatives with respect to
the components of $\vc\theta$, and $\psi$ is related to
$\kappa$ via the Poisson equation, $\nabla^2\psi=2\kappa$. The two
components of the shear, here written in complex notation, 
are derived from the traceless part of $\A$,
$\gamma(\vc\theta)= (\psi_{,11}-\psi_{,22})/2 +{\rm i}\psi_{,12}$. The
shear therefore describes the tidal part of the deflection potential
which causes the distortion of images.

Provided the density contrast $\delta$ is a homogeneous and isotropic
random field, so is the projected density $\kappa$. 
The power spectrum $P_\kappa(s)$ of $\kappa$ is related to the power
spectrum $P(\vec k)$ of the 
density fluctuations $\delta$ through
$$
P_\kappa(s) =
{9\over 4}\rund{H_0\over c}^4\Omega_{\rm d}^2
\int_0^{w_{\rm H}}\d w\;{g^2(w)\over a^2(w)} 
P\rund{{s\over f_K(w)};w}\; ;
\eqno (2)
$$
see K92 and K96 for a derivation of (2). The second argument of $P$
indicates that the power spectrum  
evolves with redshift. Several sample power spectra
$P_\kappa$ are plotted in Fig.\ts 1 of SvWJK. 

In SvWJK, the aperture mass 
$$
\m(\theta):=\int \d^2\vt\;U\rund{\abs{\vc\vt}}
\kappa(\vc\vt)
\eqno (3)
$$
was introduced as a statistics for measuring cosmic shear. Similar
quantities had previously been considered in somewhat different
contexts (e.g., Fahlman et al.\ 1994; Kaiser 1995; Kaiser et al.\ 1994;
Schneider 1996). Here,
$U(\vt)$ is a compensated filter function, i.e., 
$
\int_0^\theta \d\vt\;\vt \, U(\vt)=0
$,
which vanishes for $\vt>\theta$. The definition (3) is particularly
useful since $\m$ can directly be expressed in terms of the shear,
$$
\m(\theta)=\int \d^2\vt\;Q\rund{\abs{\vc\vt}}\,\gamma_{\rm
t}(\vc\vt)\; ,
\eqno (4)
$$
where
$
Q(\vt)=(2/ \vt^2)\int_0^\vt\d \vt'\; \vt'\,U(\vt')
-U(\vt) 
$,
and the tangential component of the shear at a position
$\vc\vt=(\vt \cos\vp,\vt\sin\vp)$ is
$
\gamma_{\rm t}(\vc\vt)=-\Re \rund{\gamma(\vc\vt)\, {\rm
e}^{-2{\rm i}\vp}}
$.
Hence, on the one hand, $\m$ yields a
spatially filtered version of the projected density field, and on the
other hand, it can be expressed simply in terms of the shear. Since in
the weak lensing regime, the observed galaxy ellipticities provide an
unbiased estimate of the local shear, $\m$ is directly related to
observables. The dispersion of $\m$ is
related to the power spectrum $P_\kappa(s)$ by
$$
\ave{\m^2(\theta)}=2\pi \int_0^\infty \d s\;s\,P_\kappa(s)
\eckk{I(s\theta)}^2\; ,
\eqno (5)
$$
where
$
I(\eta)=\int_0^1\d x\;x\,u(x)\,{\rm J}_0(\eta x)
$;
here, we have written $U(\vt)=u(\vt/\theta)/\theta^2$, and ${\rm
J}_n(x)$ denotes the Bessel function of first kind. Hence,
$\m(\theta)$ provides a 
filtered version of the projected power spectrum, and the width of the
filter, here expressed by $I^2$, depends on the choice of $u$. 

Provided that galaxies are biased tracers of the underlying (dark)
matter distribution, the expected number density of galaxies in the
direction $\vc\theta$ is given by (cf.\ Bartelmann 1995; Dolag \&
Bartelmann 1997; Sanz et al.\ 1997)
$$
N(\vc\theta)=\bar N\eck{1+b\int\d w\;p_{\rm
g}(w)\,\delta\rund{f_K(w)\vc\theta,w}}\; ,
\eqno (6)
$$
where $p_g(w)$ is the probability distribution of the galaxies in
comoving distance (or, equivalently, redshift), which depends on the
selection criteria of the galaxy sample (such as limiting magnitude,
color, etc.), $\bar N$ is the mean number density,
and $b$ is the average bias factor for this galaxy sample.
In analogy to the aperture mass, we define the aperture number
counts
$$
\N(\theta)=\int\d^2 \vt\; U(\abs{\vc\vt})\,N(\vc\vt)\; ,
\eqno (7)
$$
with the same function $U$ as in (3). 

\sec{3 Density -- shear correlations}
The correlation 
between $\m(\theta)$ and $\N(\theta)$ is measured by
$$
C(\theta):=\ave{\m(\theta)\,\N(\theta)}\;,
\eqno (8)
$$
where the angular brackets denote the ensemble average. Inserting the
explicit expressions for $\m$ and $\N$, and using the same method as
in K96 to calculate the resulting projection of the
correlator $\ave{\delta\delta}$, a few manipulations
similar to those in SvWJK yield
$$
C(\theta)
=3\pi \rund{H_0\over c}^2\Omega_{\rm d}\, b\, \bar N
\int\d w\;{p_{\rm g}(w)\,g(w)\over a(w)\,f_K(w)}
\int \d s\;s\,P\rund{{s\over f_K(w)},w}\,\eckk{I(\theta s)}^2 \;.
\eqno (3.4)
$$
Hence, the correlation $C$ depends on the cosmological model, the
redshift distributions of the galaxies which are used to estimate the
shear (which we shall call `background' galaxies in the following,
though this should not imply that all these galaxies are lying behind
those from which $\N$ is measured)
and those with which $\N$ is estimated (`foreground' galaxies). In
particular, $C$ is 
proportional to the bias factor $b$. In Sect.\ts 5 below, we shall
calculate $C$ for a simple model of the power spectrum of cosmic
density fluctuations, but first we turn to practical estimators
of $C$.

\sec{4 Practical estimators}
Assume that in a circular aperture of radius $\theta$ there are
$N_{\rm b}$ galaxies at positions $\vc\vt_i$ 
whose ellipticity is
measured and which are thus used to estimate the shear, and that
$N_{\rm f}$ galaxies with positions $\vc\vp_j$ are used for measuring
$\N$. An estimator for $\m$ is
$$
\hat\m(\theta)={\pi\theta^2\over N_{\rm b}}\sum_{i=1}^{N_{\rm b}}
Q(\abs{\vc\vt_i})\,\eps_{{\rm t}i}\;,
\eqno (10)
$$
where $\eps_{{\rm t}i}$ is the tangential component of the image
ellipticity, defined in analogy to $\gamma_{\rm t}$. In the limit of weak
lensing, to be considered here, the relation $\eps=\eps^\s+\gamma$
between image ellipticity $\eps$ and intrinsic source ellipticity
$\eps^\s$ holds, so that $\eps_i$ is an unbiased estimate of
$\gamma(\vc\vt_i)$ because the intrinsic orientations of the source
galaxies are assumed to be random. 
An estimator for $\N$ is
$$
\hat\N(\theta)=\sum_{k=1}^{N_{\rm f}}U(\abs{\vc\vp_j})\; ,
\eqno (11)
$$
so that an estimator for $C$ reads
$
\hat C(\theta)=\hat\m(\theta)\,\hat\N(\theta)
$.
To obtain the expectation value ${\rm E}(\hat C)$, several averages 
have to be taken. First, one has to average over the intrinsic
source ellipticity distribution. Denoting this operator by ${\tt A}$,
one finds
$
{\tt A}\rund{\hat C(\theta)}={\tt
A}\rund{\hat\m(\theta)}\,\hat\N(\theta)
$,
since $\hat \N$ is unaffected by ${\tt A}$. The next average has to be
taken over the galaxy positions. The corresponding operator ${\tt P}$
factorizes into two operators ${\tt P}_1$ and ${\tt P}_2$, which read
$
{\tt P}_1=\prod_{i=1}^{N_{\rm b}}\int{\d^2\vt_i \over \pi
\theta^2}$, $
{\tt P}_2=\prod_{k=1}^{N_{\rm f}}\int \d^2\vp_k\,p(\vc\vp_k)
$
where it has been ussumed that the `background' galaxies are
distributed randomly in angle, and the probability density $p(\vc\vp_k)$ for
finding a `foreground' galaxy at $\vc\vp_k$ is
$
p(\vc\vp_k)=N(\vc\vp_k)[ \int \d^2\vp\;N(\vc\vp)]^{-1}
\approx {N(\vc\vp_k)/ N_{\rm f}}$.
Thus, we neglect deviations of the total number of `foreground'
galaxies in the circular aperture from the expected number; these
deviations are of minor importance only, provided $N_{\rm f}\gg
1$. Then, by performing both averages, one finds that
the expectation value of $\hat C(\theta)$ is indeed
$C(\theta)$,
$
{\rm E}(\hat C(\theta))\equiv \langle{\tt P}({\tt A}(\hat
C(\theta)))\rangle=C(\theta)$.

We next consider the dispersion of $C$ for the case that the
`foreground' galaxies are unrelated to the matter distribution which
distorts the background galaxies, or in other words, that $\m$ and
$\N$ are uncorrelated. In that case, the expectation
value of $\hat C$ vanishes, and
$$
\sigma_0^2:= {\rm E}\rund{\hat C^2}={\rm E}\rund{\hat \m^2}{\rm
E}\rund{\hat\N^2}\;.
\eqno (12)
$$
For the first of these factors, one finds with the same
methods as used in SvWJK that
$$
{\rm E}(\hat \m^2)=\ave{\m^2(\theta)} +
{\sigma_\eps^2\,G\over 2 N_{\rm b}} \; ,
\eqno (13)
$$
where $G=\pi\theta^2\int\d^2\vt\;Q^2(\abs{\vc\vt})$, $\sigma_\eps$ is
the dispersion of the intrinsic galaxy ellipticities,
and a subdominant
`shot-noise' term has been dropped.
For the second term in (4.9), one finds
$$
{\rm E}(\hat\N^2)=\ave{\N^2(\theta)}
+{\bar N\over \pi\theta^2}\hat G\; ,
\eqno (14)
$$
where $\hat G=\pi\theta^2\int\d^2\vt\;U^2(\abs{\vc\vt})$.
The dispersion of $\N$ can be obtained either directly from observations,
or can be calculated following the biasing hypothesis. With steps
very similar to those used for deriving (9), one obtains
$$
\ave{\N^2(\theta)}=2\pi\,b^2\,\bar N^2\int\d w\;
{p_{\rm g}^2(w)\over f_K^2(w)}
\int \d s\;s\,P\rund{{s\over f_K(w)},w}
I^2(\theta s)\; .
\eqno (15)
$$
\vfill\eject
\sec{5 Analytical estimates}
This section presents 
analytical estimates for $C$ and the corresponding
signal-to-noise ratio, which are obtained
after several simplifications. First, we shall assume an Einstein-de
Sitter (EdS) cosmological model. Second, only angular scales larger than
$\sim 10$\ts arcminutes 
are considered; for an estimate on such large angular scales, the
power spectrum of the density fluctuations can be assumed to 
evolve linearly, so that $P(k,w)=a^2(w)\,P(k,0)\equiv
a^2(w)\,P_0(k)$. Third, since by a convenient choice of the function
$U(\vt)$, the resulting filter function $I^2$ is quite narrow (see
Fig.\ts 2 of SVwJK), only a small range of wavenumbers contribute to
the integrals over the power spectrum; hence, we can locally
approximate the power spectrum by a power law in $k$. And finally, we
shall choose the redshift distributions of sources and `foreground'
galaxies to be very localized.

Thus, let 
$
P_0(k)=A\, k^n
$,
with a slope $n$, and an amplitude which shall be determined from the
rms fluctuations $\sigma_8$ in a sphere of radius $R=8 h^{-1}$\ts Mpc.
Then,
$
A={2\pi^2\zeta_1^{-1}}\sigma_8^2\,R^{(3+n)}$,
with
$
\zeta_1=9\int_0^\infty\d x\; x^{n-4} [\sin(x)-x\cos(x)]^2
% =(9\sqrt{\pi}/8)\,(1+n)\Gamma({n/ 2}-{3/ 2}) /
% \Gamma(1-{n/2})
$.
For EdS, one finds 
$
f_K(w)=w=(2c/ H_0)[1-(1+z)^{-1/2}]$, $w_{\rm H}={2c/ H_0}$.
% ={2c\over H_0}\rund{1-\sqrt{a}}\quad ,
% \quad w_{\rm H}={2c\over H_0}\; .
% \eqno (5.4)
% $$
We assume that the background sources are all at the same redshift
$z_{\rm s}$, so that $p_z(z)=\delta_{\rm D}(z-z_{\rm s})$, and 
$
g(w)=(w_{\rm s}-w)/ w_{\rm s}$  for $w<w_{\rm s}$ and zero otherwise,
and $w_{\rm s}$ is the comoving distance out to 
redshift $z_{\rm s}$. Similarily, we assume that the foreground
galaxies are well localized around the redshift $z_{\rm g}$,
corresponding to comoving distance $w_{\rm g}$. 
Except for the
calculation of $\ave{\N^2}$ below, we shall approximate $p_{\rm g}$ by
a delta `function'. 

The compensated filter function $U(\vt)$ will be the same as in SvWJK,
namely\fonote{This choice corresponds to $\ell=1$ in SvWJK}
$
u(x)=(9/ \pi)(1-x^2)(1/ 3-x^2)
$
for which the corresponding function $Q(\vt)=q(\vt/\theta)/\theta^2$
is 
$
q(x)=(6/\pi)\,x^2\,(1-x^2)
$.
Then,
$
I(\eta)=(12/\pi)\,({\rm J}_4(\eta)/ \eta^2)
$,
and  $G=\hat G=6/5$.

With these assumptions and simplifications, we now estimate
$C(\theta)$ according to (9): Defining dimensionless comoving
distances by $\hat w\equiv w H_0/c$, (9) becomes
$$
% \eqalign{
C(\theta)
% &=3\pi \rund{H_0\over c}^3b\,\bar N {\hat w_{\rm s}-\hat
% w_{\rm g}\over a(w_{\rm g}) \hat w_{\rm g}\hat w_{\rm s}}
% \int\d s\;s\,a^2(w_{\rm g})\,A\rund{s\over w_{\rm g}}^n 
% I^2(\theta s)\cr
% &
={6\pi^3\zeta_2\,r^{3+n}\over \zeta_1}b\,\bar N\,\sigma_8^2\;
{a(w_{\rm g})\rund{\hat w_{\rm s}-\hat w_{\rm g}}\over \hat w_{\rm s}
\hat w_{\rm g}^{(1+n)}}\,\theta^{-(2+n)}\; ,
% \cr }
\eqno (16)
$$
where 
$
r=R H_0/c=(8/ 3)\times 10^{-3}
$
and
$
\zeta_2=\int_0^\infty \d x\;x^{1+n}\,I^2(x)
% =72  \pi^{-5/2}\,\Gamma({3/ 2}-{n/ 2})
% \Gamma(3+n/ 2)
% [\Gamma(2-{n/ 2})]^{-1}[\Gamma(6-{n/ 2})]^{-1}
$.
We shall write (16) in the following convenient form,
$$
C(\theta) = 1.822\times 10^{-3}\,\zeta(n)\, b\,\bar N\,\sigma_8^2\;
{a(w_{\rm g})\rund{\hat w_{\rm s}-\hat w_{\rm g}}\over \hat w_{\rm s}
\hat w_{\rm g}^{(1+n)}}\,\rund{\theta\over 10'}^{-(n+2)}\; ,
\eqno (17)
$$
where a good approximation of $\zeta(n)$ is 
$
\zeta(n)\approx 1+1.04(n+3/2)
+0.275 (n+3/ 2)^2 $,
which is accurate to better than $0.2\%$ for $n\in[-2,-1]$.

Next we calculate $\ave{\N^2}$ from (15) with the assumptions listed
above. However, the occurrence of the factor $p_{\rm
g}^2(w)$ in the integrand of (15) shows that we cannot simply use a
delta-function distribution of the redshifts of the `foreground'
galaxies. The reason for this problem lies is the derivation of
(15): it is valid only if the functions which are projected vary on
a much larger scale than the largest wavelength on which the power
spectrum  has
appreciable amplitude. This condition is obviously violated if the
`foreground' galaxies are very sharply localized. 
Therefore, we
assume that $p_{\rm g}$ is constant in an interval of width $\nu
w_{\rm g}$ around $w_{\rm g}$, with $\nu\ll 1$.
Then, 
$$\eqalign{
\ave{\N^2(\theta)}&={4\pi^3 \zeta_2\,r^{3+n}\over \zeta_1}\,
b^2\,\bar N^2\,\sigma_8^2\,{a^2(w_{\rm g})\over \nu\,\hat w_{\rm
g}^{n+3}}\,\theta^{-(n+2)} \cr 
&= 1.215\times 10^{-3}\, \zeta(n)\,b^2\,\bar N^2\,
\sigma_8^2\,{a^2(w_{\rm g})\over \nu\,\hat w_{\rm
g}^{n+3}}\,\rund{\theta\over 10'}^{-(n+2)}\; . \cr}
\eqno (18)
$$
We compare the two terms in (14) to see which one
dominates. Therefore, consider the ratio
$$%\eqalign{
{1\over\lambda_1}:={\ave{\N^2(\theta)}\pi\theta^2\over \bar N\,\hat G} 
% =1.012\times 10^{-3}\,\zeta(n)\,b^2\,N_{\rm f}\,
% \sigma_8^2\,{a^2(w_{\rm g})\over \nu\,\hat w_{\rm
% g}^{n+3}}\,\rund{\theta\over 10'}^{-(n+2)}\cr
% &
=1.590\, \zeta(n)\,{b^2\,\sigma_8^2\,a^2(w_{\rm g})
\over \nu\,\hat w_{\rm g}^{n+3}}
\rund{\bar N\over 5{\rm arcmin}^{-2}}\rund{\theta\over 10'}^{-n}
\;.
\eqno (19)
$$
We thus see that $\lambda_1$ is typically smaller than
unity in a situation when $\nu\ll1$ and $\hat w_{\rm g}<1$ (note that
$\hat w_{\rm g}=1/3$ corresponds to $z_{\rm g}=0.44$). 

For the same model, 
$\ave{\m^2(\theta)}$ can be calculated. One finds:
$$
\ave{\m^2(\theta)}=2.733\times 10^{-3}\,\zeta(n)\,
\sigma_8^2\,\hat w_{\rm s}^{1-n}\,
Z(n)\,\rund{\theta\over 10'}^{-(2+n)}\; ,
\eqno (20)
$$
where $Z(n)=\int_0^1\d x\;(1-x)^2\,x^{-n}=2/[ (1-n)(2-n)(3-n)]$.
Comparing the two terms in (13), we define the
ratio
$$\eqalign{
{1\over \lambda_2}&:= {\ave{\m^2(\theta)}\,2\,N_{\rm b}\over
\sigma_\eps^2\,G} \cr
&=35.8\, \zeta(n)\,\sigma_8^2\,\hat w_{\rm s}^{1-n}\,\rund{Z(n)\over 0.05}
\rund{\theta\over 10'}^{-n}\rund{n_{\rm g}\over 20\,{\rm arcmin}^{-2}}
\rund{\sigma_\eps\over 0.2}^{-2} ,  \cr }
\eqno (21)
$$
where $Z(-1.5)$ was taken as a fiducial value. We
therefore conclude that the first term in (13)
dominates in all cases of interest here (i.e., for angular scales
larger than $\sim10'$), $\lambda_2\ll 1$. 
With (18) and (20), the signal-to-noise ratio becomes
$$
{C\over\sigma_0}= \rund{1-{w_{\rm g}\over w_{\rm s}}}
\rund{w_{\rm g}\over w_{\rm s}}^{1-n\over 2}\sqrt{\nu\over Z(n)}
{1\over\sqrt{(1+\lambda_1)(1+\lambda_2)}}
\; ,
\eqno (22)
$$
with $\lambda_1<1$, $\lambda_2\ll 1$ in typical situations. This ratio
is indeed encouragingly large: Consider, for example, foreground and
background galaxies with distance ratio $w_{\rm g}/w_{\rm s}=1/2$;
then, for a width parameter $\nu=0.2$ and
spectral index $n=-1.5$, the signal-to-noise ratio is 0.42 times the
$\lambda$-dependent terms. Note that the dependence of the
signal-to-noise ratio on the number density of foreground and
background galaxies enters only through the $\lambda_i$-factors; as
long as $\lambda_i\ll 1$, $C/\sigma_0$ is independent of these
densitites.

\sec{6 Discussion}
In this letter, a statistical measure for the correlation $C$ of cosmic
shear with the number density of `foreground' galaxies was defined and
calculated in terms of the power spectrum of cosmic density
fluctuations. A practical unbiased estimator for this correlation was
defined, and its dispersion calculated. On large angular scales,
linear theory yields an accurate estimate for the power spectrum of
cosmic density fluctuations; in the framework of this approximation, 
$C$ and the corresponding signal-to-noise ratio were evaluated
explicitly.  

A measurement of $C$ would yield, for each cosmological model and
initial power spectrum $P(k)$,  
a direct estimate of the bias factor $b$
averaged over angular scale $\theta$. The method allows to probe the
scale and redshift dependence of the bias parameter.
On a short term, the detection of cosmic shear via the correlation $C$
is perhaps more
useful: 
assuming the validity of the biasing hypothesis, $C$ is a first-order
measure of the cosmic shear. The large signal-to-noise ratio (22) per
single field shows that it should be much easier to get a significant
detection of $C$ than for the previously proposed quadratic estimators
of the shear.

Taking the EIS as an example, with `foreground' galaxies
chosen to have $I\le 21$, and `background' galaxies with $22\le I\le
23.5$, the number densities will be approximately 3 and 7 per
arcmin$^2$, and the 
characteristic redshifts of the two galaxy populations will be
$\sim0.3$ and $\sim0.8$, respectively (e.g., Lilly et al.\ts 1995).
Hence, even in this case the $\lambda_i$ are smaller
than unity, and the signal-to-noise ratio per field will be of order
0.3. Thus, with $\sim 100$ fields taken from the EIS, a 3-$\sigma$
detection of the correlation on angular scales larger than $\sim 10'$
should be possible, provided the data are 
of sufficient image quality. 

In a future publication, $C$ will be calculated on smaller angular
scales, using the fully non-linear evolution of the power spectrum
(e.g., Peacock \& Dodds 1996), and for different cosmological
models. It is expected that the signal-to-noise ratio is not 
strongly affected by the non-linear evolution, and that an accurate
measurement of $C(\theta)$ on small angular scales ($\sim$ few
arcminutes) will be possible with a moderate amount of high-quality
image data.

I would like to thank  L.\ da Costa, E.\ Mart\'\i
nez-Gonz\'alez \& L.\ van Waerbeke for fruitful
discussions, and M.\ Bartelmann \& S.\ Seitz in addition for helpful
comments on the manuscript. \SFB 

\vfill\eject
\def\ref#1{\vskip1pt\noindent\hangindent=40pt\hangafter=1 {#1}\par}
\sec{References}

\ref{Bartelmann, M.\ 1995, A\&A 298, 661}
% \ref{Bartelmann, M. \& Schneider, P.\ 1991, A\&A 248, 349}
% \ref{Bartelmann, M. \& Schneider, P.\ 1994, A\&A 284, 1.}
% \ref{Baugh, C., Gazta\~naga, E. \& Efstathiou, G. \ 1995, MNRAS 274, 1049}
% \ref{Ben\'\i tez, N. \& Mart\'\i nez-Gonz\'alez, E.\ 1997,
% ApJ 477, 27}
% \ref{Bernardeau, F., van Waerbeke, L. \& Mellier, Y.\ 1997, A\&A 322,
% 1 (BvWM)}
% \ref{Blandford, R.D. \& Jaroszy\'nski, M.\ 1981, ApJ 246, 1}
\ref{Blandford, R.D., Saust, A.B., Brainerd, T.G. \& Villumsen, J.V.\
1991, MNRAS 251, 600}
% \ref{Bonnet, H. \& Mellier, Y.\ 1995, A\&A 303, 331.}
% \ref{Bouchet, F., Juszkiewicz, R., Colombi, S. \& Pellat, R.\ 1992, ApJ 394, L5.}
% \ref{Bower, R. \& Smail, I.\ 1997, astro-ph/9612151} 
% \ref{Colombi, S., Bouchet F.R. \& Hernquist L.\ 1996, ApJ, 465, 14}
\ref{Dolag, K. \& Bartelmann, M.\ 1997, MNRAS, in press, also: astro-ph/9704217}
% \ref{Efstathiou, G.\ 1996, in: {\it Cosmology and large scale
% structure}, Les Houches Session LX, R. Schaeffer, J. Silk, M. Spiro \&
% J. Zinn-Justin (eds.), North-Holland, p. 133.}
\ref{Fahlman, G., Kaiser, N., Squires, G. \& Woods, D.\ 1994, ApJ 437, 56.}
% \ref{Fort, B., Mellier, Y., Dantel-Fort, M., Bonnet, H. \& Kneib,
% J.-P.\ 1996, A\&A 310, 705 }
% \ref{Fry, J.N. \ 1984, ApJ 279, 499.}
% \ref{Gazta\~naga, E. \& Bernardeau, F.\ 1997, astro-ph/9707095}
% \ref{Goroff, M.H., Grinstein, B., Rey, S.J. \& Wise, M.B.\ 1986, MNRAS 236, 385.}
% \ref{Gunn, J.E.\ 1967, ApJ 147, 61}
% \ref{Hamilton, A.J.S., Kumar, P., Lu, E. \& Matthews, A.\ 1991, ApJ
% 374, L1}
% \ref{Jain, B., Mo, H. \& White, S.D.M.\ 1995, MNRAS 276, L25}
\ref{Jain, B. \& Seljak, U.\ 1997, ApJ 484, 560}
% \ref{Jaroszy\'nski, M.\ 1991, MNRAS 249, 430} 
% \ref{Jaroszy\'nski, M.\ 1992, MNRAS 255, 655} 
% \ref{Jaroszy\'nski, M., Park, C., Paczy\'nski, B. \& Gott, J.R.\ 1990,
% ApJ 365, 22}
\ref{Kaiser, N.\ 1992, ApJ 388, 272 (K92)}
% \ref{Kaiser, N.\ 1995, ApJ 439, L1}
\ref{Kaiser, N.\ 1996, astro-ph/9610120 (K96)}
% \ref{Kaiser, N., Squires, G. \& Broadhurst, T.\ 1995, ApJ 449, 460.}
\ref{Kaiser, N., Squires, G., Fahlman, G. \& Woods, D.\ 1994, in: {\it
Clusters of Galaxies}, eds. F.\ts Durret, A.\ts Mazure \& J.\ts Tran
Thanh Van, Editions Frontieres.}
% \ref{Lee, M.H.\ \& Paczy\'nski, B.\ 1990, ApJ 357, 32}
\ref{Lilly, S.J., Tresse, L., Hammer, F., Crampton, D. \& Le F\`evre,
O.\ 1995, ApJ 455, 108.} 
% \ref{Luppino, G. \& Kaiser, N.\ 1997, ApJ 475, 20.}
\ref{Miralda-Escud\'e, J.\ 1991, ApJ 380, 1.}
\ref{Moessner, R., Jain, B. \& Villumsen, J.\ 1997, MNRAS, in press}
% \ref{Mould, J.\ et al.\ 1994, MNRAS 271, 31.}
\ref{Peacock, J.A. \& Dodds, S.J.\ 1996, MNRAS 280, L19}
\ref{Renzini, A. \& da Costa, L.N.\ 1997, ESO Messenger 87, 23.}
\ref{Sanz, J.L., Mart\'\i nez-Gonz\'alez, E. \& Ben\'\i tez, N.\ 1997,
MNRAS, in press, also: astro-ph/9706278}
\ref{Schneider, P.\ 1996, MNRAS 283, 837}
\ref{Schneider, P., van Waerbeke, L., Jain, B. \& Kruse, G.\ 1997a,
astro-ph/9708143 (SvWJK)}
\ref{Schneider, P., van Waerbeke, L., Mellier, Y., Jain, B., Seitz,
S. \& Fort, B.\ 1997b, astro-ph/9705122}
% \ref{Seitz, S., Schneider, P. \& Ehlers, J.\ 1994,
% Class. Quant. Grav. 11, 2345}
% \ref{Smail, I., Hogg, D.W., Yan, L. \& Cohen, J.G. 1995, ApJ, 449, L105}
\ref{van Waerbeke, L., Mellier, Y., Schneider, P., Fort, B. \& Mathez, G.
\ 1997, A\& A, 317, 303.}
% \ref{Villumsen, J.\ 1995, astro-ph/9507007}
\ref{Villumsen, J.\ 1996, MNRAS 281, 369.}
% \ref{Wambsganss, J., Cen, R., Ostriker, J.P. \& Turner, E.L.\ 1995,
% Science 268, 274}
% \ref{Wambsganss, J., Cen, R., Xu, G. \& Ostriker, J.P.\ 1997, ApJ 475,
% L81}

\vfill\eject\end